\documentclass[preprint,prd,aps,showpacs,showkeys,nofootinbib]{revtex4}
\usepackage{amssymb}
\usepackage{amsmath}
\allowdisplaybreaks[4]
\usepackage{multirow}
\usepackage{graphicx}
\usepackage{float}

\begin{document}

\preprint{}

\title{Searching for lepton flavor violating decays $\tau\rightarrow Pl$ in minimal R-symmetric supersymmetric standard model}

\author{Ke-Sheng Sun$^a$\footnote{sunkesheng@126.com;sunkesheng@bdu.edu.cn},
Tao Guo$^{b}$\footnote{sjzueguotao@163.com},
Wei Li$^{c,d}$\footnote{watliwei@163.com},
Xiu-Yi Yang$^e$\footnote{yxyruxi@163.com},
Shu-Min Zhao$^{c,d}$\footnote{zhaosm@hbu.edu.cn}}

\affiliation{$^a$Department of Physics, Baoding University, Baoding, 071000,China\\
$^b$School of Mathematics and Science, Hebei GEO University, Shijiazhuang, 050031, China\\
$^c$Department of Physics, Hebei University, Baoding, 071002, China\\
$^d$Key Laboratory of High-Precision Computation and Application of
Quantum Field Theory of Hebei Province, Baoding, 071002, China\\
$^e$School of science, University of Science and Technology Liaoning,
Anshan, 114051, China}

\begin{abstract}

We analyze the lepton flavor violating decays $\tau\rightarrow Pl$ ($P=\pi,\eta,\eta';\;l=e,\mu$) in the scenario of the minimal R-symmetric supersymmetric standard model. The prediction on the branching ratios BR$(\tau\rightarrow P e)$ and BR$(\tau\rightarrow P \mu)$ is affected by the mass insertion parameters $\delta^{13}$ and $\delta^{23}$, respectively. These parameters are constrained by the experimental bounds on the branching ratios BR($\tau\rightarrow e (\mu) \gamma$) and BR($\tau\rightarrow 3e(\mu)$). The result shows $Z$ penguin dominates the prediction on BR($\tau\rightarrow Pl$) in a large region of the parameter space. The branching ratios for BR($\tau\rightarrow Pl$) are predicted to be, at least, five orders of magnitude smaller than present experimental bounds and three orders of magnitude smaller than future experimental sensitivities.

\end{abstract}



\maketitle

\section{Introduction\label{sec1}}
\indent\indent

Searching for lepton flavor violating (LFV) decays is of great importance in probing new physics (NP) beyond the standard model (SM) since the theoretical prediction on these LFV decays is suppressed by small mass of neutrinos in the SM. Much effort has been devoted to searching for LFV decays in experiment and the usually discussed decay channels are $l_2\rightarrow l_1 \gamma$, $l_2\rightarrow 3 l_1$, $\mu$-e conversion in nuclei, semileptonic $\tau$ decays, and so on. The experimental observation of LFV decays of $\tau$ lepton is one goal of a bunch of excellent dedicated experiments. The first generation of B-factories, that stand for $\tau$ factories too, like BaBar or Belle, have joined in the pursuit of charged LFV decays coming from $\tau$ lepton \cite{Cei}. All experiments have provided excellent bounds on the hadronic decays of $\tau$, for the first time \cite{Hayasaka,Belle,Belle1}, such as $\tau\rightarrow \mu (P,V,PP)$, where $P(V)$ stands for a pseudoscalar (vector) meson. The study of LFV decays of $\tau$ lepton are also one of main goals of the future SuperKEKB/Belle II project under construction at KEK \cite{Belle2}. The present upper bounds on the branching ratios of $\tau\rightarrow Pl$ ($P=\pi,\eta,\eta';\;l=e,\mu$) are shown in TABLE.\ref{TPL} \cite{PDG}.
\begin{table}[h]
\caption{Current limits on LFV decays $\tau\rightarrow Pl$. }
\begin{tabular}{@{}cccccc@{}} \toprule
Decay&Bound&Experiment&Decay&Bound&Experiment\\
\colrule
$\tau\rightarrow e\pi$&$8.0\times 10^{-8}$&BELLE \cite{BELLEt}&$\tau\rightarrow \mu\pi$&$1.1\times 10^{-7}$&BABAR \cite{BABARt}\\
$\tau\rightarrow e\eta$&$9.2\times 10^{-8}$&BELLE \cite{BELLEt}&$\tau\rightarrow \mu\eta$&$6.5\times 10^{-8}$&BELLE \cite{BELLEt}\\
$\tau\rightarrow e\eta'$&$1.6\times 10^{-7}$&BELLE \cite{BELLEt}&$\tau\rightarrow \mu\eta'$&$1.3\times 10^{-7}$&BELLE \cite{BELLEt}\\
\botrule
\end{tabular}
\label{TPL}
\end{table}
Assuming the integrated luminosity of 50 $\textup{ab}^{-1}$, the future prospects of BR$(\tau\rightarrow Pl)$ in Belle II will be extrapolated at the level of $\mathcal{O}(10^{-9}$-$10^{-10})$ \cite{Altmannshofer}.

In various extensions of the SM, corrections to BR$(\tau\rightarrow Pl)$ are enhanced by different LFV sources. There are a few studies within non-SUSY models, such as two Higgs doublet models \cite{Li,Kanemura}, 331 model \cite{hua}, TC2 models \cite{yue}, littlest Higgs model with T parity \cite{Goto}, simplest little Higgs model \cite{Lami}, leptoquark models \cite{Carpentier, Dorsner} and unparticle model \cite{LiZH}. Some models with heavy Dirac/Majorana neutrinos can have BR$(\tau\rightarrow Pl)$ close to the experimental sensitivity \cite{Garcia, Ilakovac, Ilakovac3}. In Type III seesaw model, there are tree level flavor changing neutral currents in the lepton sector which can enhance the prediction on BR$(\tau\rightarrow Pl)$ \cite{He1,Arhrib}. There are also a few studies within SUSY models, such as MSSM \cite{Fukuyama,Sher}, unconstrained MSSM \cite{brignole}, supersymmetric seesaw mechanism model \cite{Chen}, R-parity violating SUSY \cite{Saha}, the CMSSM-seesaw and NUHM-seesaw \cite{Arganda}. Within an effective field theory framework, LFV decays $\tau\rightarrow Pl$ are studied to set constraints on the Wilson coefficients of the LFV operators \cite{Black, cirigliano, He, Dorsner, Cai, Cai1, Gabrielli, Petrov, Celis, Buchmuller, Grzadkowski}.

In this paper, we will study the LFV decays $\tau\rightarrow Pl$ in the minimal R-symmetric supersymmetric standard model (MRSSM) \cite{Kribs}. The MRSSM has an unbroken global $U(1)_R$ symmetry and provides a new solution to the supersymmetric flavor problem that exists in the MSSM. In this model, R-symmetry forbids the Majorana gaugino masses, $\mu$ term, $A$ terms and all left-right squark and slepton mass mixings. The $R$-charged Higgs $SU(2)_L$ doublets $\hat{R}_u$ and $\hat{R}_d$ are introduced in the MRSSM to yield Dirac mass terms of higgsinos. The additional superfields $\hat{S}$, $\hat{T}$ and $\hat{O}$ are introduced to yield Dirac mass terms of gauginos. The most unusual characteristic in the MRSSM is that  large flavor violation is allowed in the squark and slepton mass matrices. The presence of large flavor violation in the MRSSM means that it is no longer appropriate to discuss stops or selectrons necessarily. The large flavor violation opens the possibility for a wide variety of new signals at the LHC and is worthy of significant study. Studies on phenomenology in the MRSSM can be found in Refs. \cite{Die1, Die2, Die3, Die4, Die5,Die6,KSS, Kumar, Blechman, Kribs1, Frugiuele, Jan, Chakraborty, Braathen, Athron, Alvarado,sks1,sks2}. It is interesting to explore whether BR$(\tau\rightarrow Pl)$ can be enhanced to be close to the current experiment limits or future experimental sensitivities while the predictions on other LFV processes do not exceed current experiment constraints. Thus, we choose decay channels $\tau\rightarrow Pl$ as an object of the analysis. Similar to the case in the MSSM, LFV decays mainly originate from the off-diagonal entries in slepton mass matrices $m_l^2$ and $m_r^2$. Taking into account the experimental constraints from decay channels $\tau\rightarrow l\gamma$ and $\tau\rightarrow 3l$ on the off-diagonal parameters, we investigate the branching ratios BR($\tau\rightarrow Pl$) as a function of the off-diagonal parameters and other model parameters.

The outline of this paper is organized as follows. In Section \ref{sec2}, we provide a brief introduction on the MRSSM and present the definitions of the sneutrino mass matrix and slepton mass matrix. Then, we present conventions for the effective operators and the corresponding Wilson coefficients. The existing constraints, benchmark points and the results of our calculation are shown in Section \ref{sec3}. In Section \ref{sec4}, the conclusion is drawn. The definitions of mass matrices of scalar and pseudo-scalar Higgs boson, neutralino, $\chi^{\pm}$-chargino and squarks are listed in Appendix \ref{secapp}.

\section{MRSSM\label{sec2}}

In this section, we firstly provide a simple overview of the MRSSM in order to fix the notations used in this work. The MRSSM has the same gauge symmetry $SU(3)_C\times SU(2)_L\times U(1)_Y$ as the SM and the MSSM. The spectrum of fields in the MRSSM contains the standard MSSM matter, the Higgs and gauge superfields augmented by the chiral adjoints $\hat{\cal O},\hat{T},\hat{S}$ and two $R$-Higgs iso-doublets. The superfields with R-charge in the MRSSM are given in TABLE.\ref{Field}.
\begin{table}[!htbp]
\centering
\caption{The superfields with R-charge in MRSSM. }
\begin{tabular}{|c|c|c|c|c|c|c|}
\hline
\multicolumn{1}{|c}{Field} & \multicolumn{2}{|c}{Superfield} &
                              \multicolumn{2}{|c}{Boson} &
                              \multicolumn{2}{|c|}{Fermion} \\
\hline
Gauge vector & $\hat{g},\hat{W},\hat{B}$&0& $g,W,B$&0& $\tilde{g},\tilde{W}\tilde{B}$&  +1 \\\hline
\multirow{2}*{Matter}& $\hat{l}, \hat{e}^c$& +1& $\tilde{l},\tilde{e}^*_R$&+1& $l,e^*_R$& 0   \\
 \cline{2-7}     & $\hat{q},{\hat{d}^c},{\hat{u}^c}$& +1& $\tilde{q},{\tilde{d}}^*_R,{\tilde{u}}^*_R$ & +1& $q,d^*_R,u^*_R$                             & 0 \\\hline
$H$-Higgs & ${\hat{H}}_{d,u}$&0& $H_{d,u}$& 0& ${\tilde{H}}_{d,u}$&-1 \\ \hline
R-Higgs & ${\hat{R}}_{d,u}$ & +2& $R_{d,u}$& +2& ${\tilde{R}}_{d,u}$& +1 \\\hline
Adjoint chiral& $\hat{\cal O},\hat{T},\hat{S}$&0& $O,T,S$&0& $\tilde{O},\tilde{T},\tilde{S}$ &-1 \\
\hline
\end{tabular}\label{Field}
\end{table}
The general form of the superpotential of the MRSSM is given by \cite{Die1}
\begin{equation}
\begin{array}{l}
\mathcal{W}_{MRSSM} = \mu_d(\hat{R}_d\hat{H}_d)+\mu_u(\hat{R}_u\hat{H}_u)+\Lambda_d(\hat{R}_d\hat{T})\hat{H}_d+\Lambda_u(\hat{R}_u\hat{T})\hat{H}_u
+\lambda_d\hat{S}(\hat{R}_d\hat{H}_d)\\
\hspace{5.5em}+\lambda_u\hat{S}(\hat{R}_u\hat{H}_u)-Y_d\hat{d}(\hat{q}\hat{H}_d)-Y_e\hat{e}(\hat{l}\hat{H}_d)+Y_u\hat{u}(\hat{q}\hat{H}_u),
\end{array}\label{sptl}
\end{equation}
where $\hat{H}_u$ and $\hat{H}_d$ are the MSSM-like Higgs weak iso-doublets, $\hat{R}_u$ and $\hat{R}_d$ are the $R$-charged Higgs $SU(2)_L$ doublets.  The corresponding Dirac higgsino mass parameters are denoted as $\mu_u$ and $\mu_d$. Although R-symmetry forbids the $\mu$ terms of the MSSM, the bilinear combinations of the normal Higgs $SU(2)_L$ doublets $\hat{H}_u$ and $\hat{H}_d$ with the Higgs $SU(2)_L$ doublets $\hat{R}_u$ and $\hat{R}_d$ are allowed in Eq.(\ref{sptl}). The parameters $\lambda_u$, $\lambda_d$, $\Lambda_u$ and $\Lambda_d$ are Yukawa-like trilinear terms involving the singlet $\hat{S}$ and the triplet $\hat{T}$.

For the phenomenological studies we take the soft-breaking scalar mass terms \cite{Die3}
\begin{equation}
\begin{array}{l}
V_{SB,S} = m^2_{H_d}(|H^0_d|^2+|H^{-}_d|^2)+m^2_{H_u}(|H^0_u|^2+|H^{+}_u|^2)+(B_{\mu}(H^-_dH^+_u-H^0_dH^0_u)+h.c.)\\
\hspace{3.5em}+m^2_{R_d}(|R^0_d|^2+|R^{+}_d|^2)+m^2_{R_u}(|R^0_u|^2+|R^{-}_u|^2)+m^2_T(|T^0|^2+|T^-|^2+|T^+|^2)\\
\hspace{3.5em}+m^2_S|S|^2+ m^2_O|O^2|+\tilde{d}^*_{L,i} m_{q,{i j}}^{2} \tilde{d}_{L,j} +\tilde{d}^*_{R,i} m_{d,{i j}}^{2} \tilde{d}_{R,j}+\tilde{u}^*_{L,i}  m_{q,{i j}}^{2} \tilde{u}_{L,j}\\
\hspace{3.5em}+\tilde{u}^*_{R,i}  m_{u,{i j}}^{2} \tilde{u}_{R,j}+\tilde{e}^*_{L,i} m_{l,{i j}}^{2} \tilde{e}_{L,j}+\tilde{e}^*_{R,{i}} m_{r,{i j}}^{2} \tilde{e}_{R,{j}} +\tilde{\nu}^*_{L,i} m_{l,{i j}}^{2} \tilde{\nu}_{L,j}.
\end{array}\label{soft}
\end{equation}
All the trilinear scalar couplings involving Higgs bosons to squark and slepton are forbidden in Eq.(\ref{soft}) because the sfermions have an R-charge and these terms are non R-invariant, and this has relaxed the flavor problem of the MSSM \cite{Kribs}. The Dirac nature is a manifest feature of the MRSSM fermions. The soft-breaking Dirac mass terms of the singlet $\hat{S}$, triplet $\hat{T}$ and octet $\hat{O}$ take the form as
\begin{equation}
V_{SB,DG}=M^B_D\tilde{B}\tilde{S}+M^W_D\tilde{W}^a\tilde{T}^a+M^O_D\tilde{g}\tilde{O}+h.c.,
\label{}
\end{equation}
where $\tilde{B}$, $\tilde{W}$ and $\tilde{g}$ are the usually MSSM Weyl fermions. The R-Higgs bosons do not develop vacuum expectation values (VEVs) since they carry R-charge 2.
After electroweak symmetry breaking, the singlet and triplet VEVs effectively modify the $\mu_u$ and $\mu_d$, and the modified $\mu_i$ parameters are given by
\begin{align}
\mu_d^{eff,+}= \frac{1}{2} \Lambda_d v_T  + \frac{1}{\sqrt{2}} \lambda_d v_S  + \mu_d ,\;\;
\mu_u^{eff,-}= -\frac{1}{2} \Lambda_u v_T  + \frac{1}{\sqrt{2}} \lambda_u v_S  + \mu_u.\nonumber
\end{align}
$v_T$ and $v_S$ are vacuum expectation values of $\hat{T}$ and $\hat{S}$.

There are four complex neutral scalar fields and they can mix. Assuming the vacuum expectation values are real, the real and imaginary components in four complex neutral scalar fields do not mix, and the mass-square matrix breaks into two $4\times 4$ sub-matrices. In the scalar sector all fields mix and the SM-like Higgs boson is dominantly given by the up-type field. In the pseudo-scalar sector there is no mixing between the MSSM-like states and the singlet-triplet states, and the $4\times 4$ mass-squared matrix breaks into two $2\times 2$ submatrices. The number of neutralino degrees of freedom in the MRSSM is doubled compared to the MSSM as the neutralinos are Dirac-type. The number of chargino degrees of freedom in the MRSSM is also doubled compared to the MSSM and these charginos can be grouped to two separated chargino sectors according to their R-charge. The $\chi^{\pm}$-chargino sector has R-charge 1 electric charge; the $\rho$-chargino sector has R-charge -1 electric charge. Here, we do not discuss the $\rho$-chargino sector in detail since it does not contribute to the LFV decays. More information about the $\rho$-chargino can be found in Ref.\cite{Die3,Die5,sks1,KSS}. For convenience, we present the tree-level mass matrices for scalar and pseudo-scalar Higgs bosons, neutralinos, charginos and squarks of the MRSSM in Appendix \ref{secapp}.

\begin{figure}[htbp]
\setlength{\unitlength}{1mm}
\centering
\begin{minipage}[c]{1\textwidth}
\includegraphics[width=5.5in]{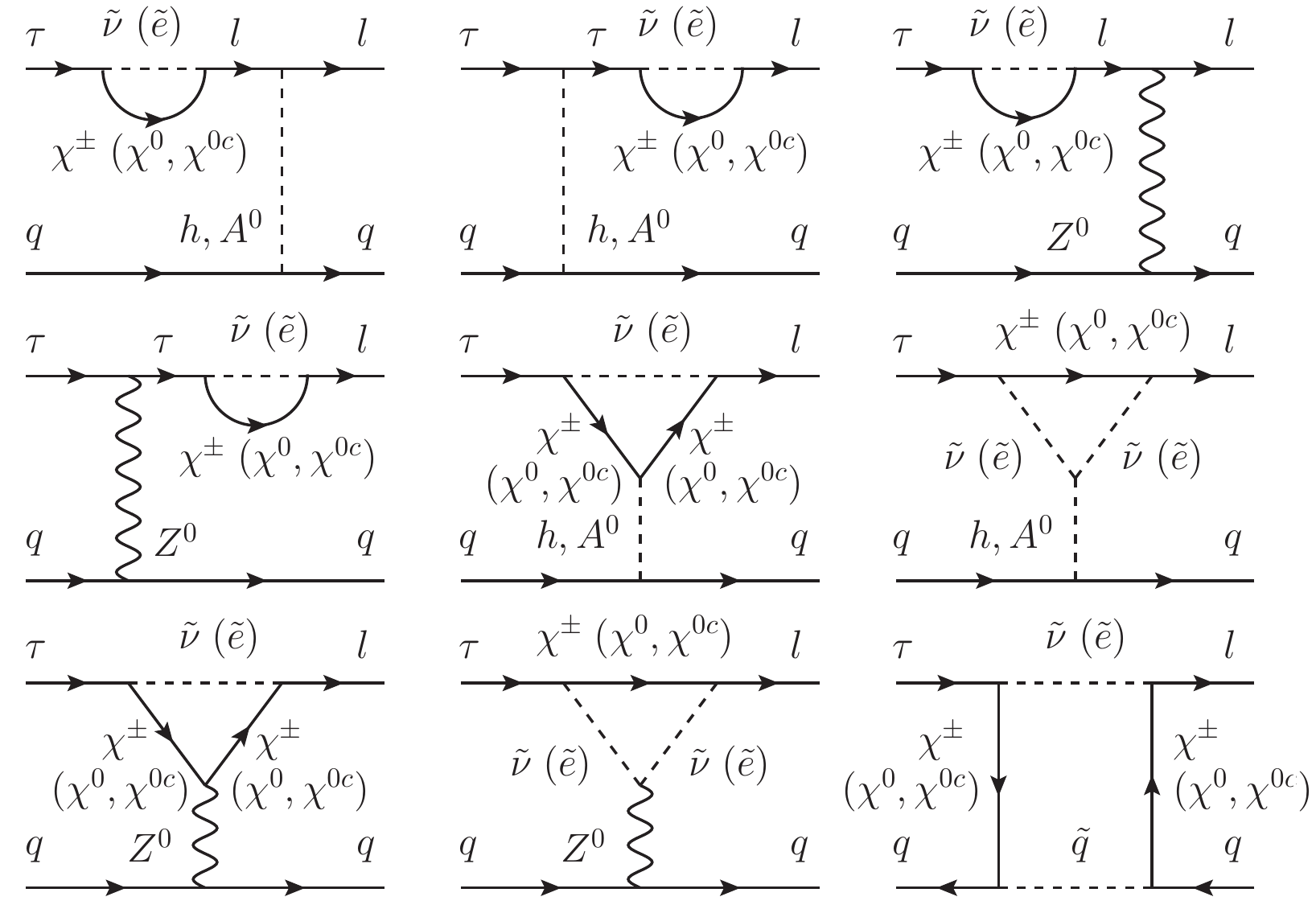}
\end{minipage}
\caption[]{Feynman diagrams contributing to $\tau\rightarrow Pl$ in the MRSSM. Corrections from crossed diagrams of box diagram are also considered.}
\label{mu-e}
\end{figure}

In MRSSM, LFV decays mainly originate from the potential misalignment in slepton mass matrices. In the gauge eigenstate basis $\tilde{\nu}_{iL}$, the sneutrino mass matrix and the diagonalization procedure are
\begin{equation}
m^2_{\tilde{\nu}} =
m_l^2+\frac{1}{8}(g_1^2+g_2^2)( v_{d}^{2}- v_{u}^{2})+g_2 v_T M^{W}_D-g_1 v_S M^{B}_D,
Z^V m^2_{\tilde{\nu}} (Z^{V})^{\dagger} = m^{2,\textup{diag}}_{\tilde{\nu}},\label{sn}
\end{equation}
where the last two terms in mass matrix are newly introduced by MRSSM. The slepton mass matrix and the diagonalization procedure are
\begin{equation}
m^2_{\tilde{e}} = \left(
\begin{array}{cc}
(m^2_{\tilde{e}})_{LL} &0 \\
0  &(m^2_{\tilde{e}})_{RR}\end{array}
\right),Z^E m^2_{\tilde{e}} (Z^{E})^{\dagger} =m^{2,\textup{diag}}_{\tilde{e}},\label{sl}
 \end{equation}
where
\begin{align}
(m^2_{\tilde{e}})_{LL} &=m_l^2+ \frac{1}{2} v_{d}^{2} |Y_{e}|^2 +\frac{1}{8}(g_1^2-g_2^2)(v_{d}^{2}- v_{u}^{2}) -g_1 v_S M_D^B-g_2v_TM_D^W ,\nonumber\\
(m^2_{\tilde{e}})_{RR} &= m_r^2+\frac{1}{2}v_d^2|Y_e|^2+\frac{1}{4}g_1^2( v_{u}^{2}- v_{d}^{2})+2g_1v_SM_D^B.\nonumber
\end{align}
The sources of LFV are the off-diagonal entries of the $3\times 3$ soft supersymmetry breaking matrices $m_l^2$ and $m_r^2$ in Eqs.(\ref{sn}, \ref{sl}). From Eq.(\ref{sl}) we can see that the left-right slepton mass mixing is absent in the MRSSM, whereas the $A$ terms are present in the MSSM. The relevant Feynman diagrams contributing to $\tau\rightarrow Pl$ are presented in FIG.\ref{mu-e}.

We now focus on the LFV processes $\tau\rightarrow Pl$. Using the effective Lagrangian method, we present the analytical expression for the decay width of $\tau\rightarrow Pl$. At the quark level, the interaction Lagrangian for $\tau\rightarrow Pl$ can be written as \cite{Flavor}
\begin{equation}
{\cal L}_{\tau\rightarrow Pl}=\sum_{I=S,V}^{X,Y=L,R}B^I_{XY}(\bar{l}_{\beta}\Gamma_IP_X\tau)(\bar{d}\Gamma_IP_Yd)
+C^I_{XY}(\bar{l}_{\beta}\Gamma_IP_X\tau)(\bar{u}\Gamma_IP_Yu)+h.c.,
\label{BC}
\end{equation}
where the index $\beta$(=$1,2$) denotes the generation of the emitted lepton and $l_1(l_2)=e(\mu)$. Since only the axial-vector current contributes to $\tau\rightarrow Pl$, the coefficients in Eq.(\ref{BC}) do not include photon penguin contribution but they include $Z$ boson and scalar ones. The contribution to the Wilson coefficients $C^I_{XY}$ and $B^I_{XY}$ in Eq.(\ref{BC}) can be classified into self-energies, $Z $ penguins, Higgs penguins and box diagrams, as shown in FIG.\ref{mu-e}. Now consider the implication of virtual Higgs exchange for $\tau\rightarrow Pl$. Both the contributions from scalar and pseudo-scalar Higgs sector are considered in this work. However, all the Higgs penguin contribution is negligible since the couplings of scalar and pseudo-scalar Higgs to the light quarks are suppressed by their masses.

Then the decay width for $\tau\rightarrow Pl$ is given by
\begin{equation}
\Gamma(\tau\rightarrow Pl)=\frac{\lambda^{1/2}(m_{\tau}^2,m^2_l,m^2_P)}{16\pi m_{\tau}^3}\sum_{i,f}|{\cal M}|^2,
\end{equation}
where the averaged squared amplitude can be written as
\begin{equation}
\sum_{i,f}|{\cal M}|^2=\sum_{I,J=S,V}[2m_{\tau}m_l(a^I_Pa^{J\ast}_P-b^I_Pb^{J\ast}_P)+(m_\tau^2+m^2_l-m_P^2)(a^I_Pa^{J\ast}_P
+b^I_Pb^{J\ast}_P).
\end{equation}
The coefficients $a^{S,V}_P$ and $b^{S,V}_P$ are linear combinations of the Wilson coefficients in Eq.(\ref{BC}),
\begin{eqnarray}
a^S_P&=&\frac{f_\pi}{2}\sum_{X=L,R}[\frac{D^d_X(P)}{m_d}(B^S_{LX}+B^S_{RX})+\frac{D^u_X(P)}{m_u}(C^S_{LX}+C^S_{RX})],\nonumber\\
b^S_P&=&\frac{f_\pi}{2}\sum_{X=L,R}[\frac{D^d_X(P)}{m_d}(B^S_{RX}-B^S_{LX})+\frac{D^u_X(P)}{m_u}(C^S_{RX}-C^S_{LX})],\nonumber\\
a^V_P&=&\frac{f_\pi}{4}C(P)(m_\tau-m_l)[-B^V_{LL}+B^V_{LR}-B^V_{RL}+B^V_{RR}+C^V_{LL}-C^V_{LR}+C^V_{RL}-C^V_{RR}],\nonumber\\
b^V_P&=&\frac{f_\pi}{4}C(P)(m_\tau+m_l)[-B^V_{LL}+B^V_{LR}+B^V_{RL}-B^V_{RR}+C^V_{LL}-C^V_{LR}-C^V_{RL}+C^V_{RR}],\nonumber
\end{eqnarray}
where $f_\pi$ is the pion decay constant. The expressions for coefficients $C(P)$,$D^{d,u}_{L}(P)$ are listed in TABLE.\ref{Pco} \cite{Arganda}.
\begin{table}[h]
\caption{Coefficients for each pseudoscalar meson $P$ }
\begin{tabular}{@{}cccc@{}} \toprule
&$P=\pi$&$P=\eta$&$P=\eta'$\\
\colrule
$ C(P)$&1&$\frac{1}{\sqrt{6}}(\sin\theta_\eta+\sqrt{2}\cos\theta_\eta)$&$\frac{1}{\sqrt{6}}(\sqrt{2}\sin\theta_\eta-\cos\theta_\eta)$\\
$D^d_L(P)$&-$\frac{m^2_\pi}{4}$&$\frac{1}{4\sqrt{3}}[(3m^2_\pi-4m^2_K)\cos\theta_\eta-2\sqrt{2}m^2_K\sin\theta_\eta]$
&$\frac{1}{4\sqrt{3}}[(3m^2_\pi-4m^2_K)\sin\theta_\eta+2\sqrt{2}m^2_K\cos\theta_\eta]$\\
$D^u_L(P)$&$\frac{m^2_\pi}{4}$&$\frac{1}{4\sqrt{3}}m^2_\pi(\cos\theta_\eta-\sqrt{2}\sin\theta_\eta)$
&$\frac{1}{4\sqrt{3}}m^2_\pi(\sin\theta_\eta+\sqrt{2}\cos\theta_\eta)$\\
\botrule
\end{tabular}
\label{Pco}
\end{table}
Here, $m_{\pi}$ and $m_K$ denote the masses of the neutral pion and Kaon, and $\theta_{\eta}$ denotes the $\eta-\eta'$ mixing angle. In addition, $D^{d,u}_{R}(P)$ = - $(D^{d,u}_{L}(P))^{\ast}$.

Finally, the MRSSM has been implemented in the Mathematica package SARAH-4.14.3 \cite{SARAH, SARAH1, SARAH2,Flavor}. The masses of the MRSSM particles, mixing matrices and the Wilson coefficients of the corresponding operators in the effective lagrangian are computed by SPheno-4.0.4 \cite{SPheno1, SPheno2} modules written by SARAH.

\section{Numerical Analysis\label{sec3}}
\indent\indent
The calculations of BR($\tau\rightarrow Pl$) in the MRSSM are evaluated within the framework of SARAH-4.14.3 \cite{SARAH, SARAH1, SARAH2,Flavor} and SPheno-4.0.4 \cite{SPheno1,SPheno2}. The experimental values of the Higgs mass and the W boson mass can impose stringent and nontrivial constraints on the model parameters. The one loop and leading two loop corrections to the lightest (SM-like) Higgs boson in the MRSSM have been computed in \cite{Die3} and the new fields and couplings can give large contributions to the Higgs mass even for stop masses of order 1 TeV and no stop mixing. Meanwhile, the new fields and couplings can not give too large contribution to the W boson mass and muon decay in the same regions of parameter space. A better agreement with the latest experimental value for the W boson mass has been  investigated in \cite{Die6}. It combines all numerically relevant contributions that are known in the SM in a consistent way with all the MRSSM one loop corrections. A set of the updated benchmark point BMP1 is given in \cite{Die6} and we display them in Eq.(\ref{N1}) where all the mass parameters are in GeV or GeV$^2$.
\begin{equation}
\begin{array}{l}
\tan\beta=3,B_\mu=500^2,\lambda_d=1.0,\lambda_u=-0.8,\Lambda_d=-1.2,\Lambda_u=-1.1,
\\
M_D^B=550,M_D^W=600,\mu_d=\mu_u=500,v_S=5.9,v_T=-0.38,\\
(m^2_l)_{11}=(m^2_l)_{22}=(m^2_l)_{33}=(m^2_r)_{11}
=(m^2_r)_{22}=(m^2_r)_{33}=1000^2,\\
(m^2_{\tilde{q}})_{11}=(m^2_{\tilde{u}})_{11}=(m^2_{\tilde{d}})_{11}=(m^2_{\tilde{q}})_{22}
=(m^2_{\tilde{u}})_{22}=(m^2_{\tilde{d}})_{22}=2500^2,\\
(m^2_{\tilde{q}})_{33}=(m^2_{\tilde{u}})_{33}=(m^2_{\tilde{d}})_{33}=1000^2,m_T=3000,m_S=2000.
\end{array}\label{N1}
\end{equation}
In the numerical analysis, the default values of the input parameters are set same with those in Eq.(\ref{N1}). The off-diagonal entries of squark mass matrices $m^2_{\tilde{q}}$, $m^2_{\tilde{u}}$, $m^2_{\tilde{d}}$ and slepton mass matrices $m^2_l$, $m^2_r$ in Eq.(\ref{N1}) are zero. The large value of $|v_T|$ is excluded by measurement of the W boson mass because the VEV $v_T$ of the $SU(2)_L$ triplet field $T^0$ gives a correction to the W mass through \cite{Die1}
\begin{align}
m_W^2=\frac{1}{4}g_2^2(v_u^2+v_d^2)+g_2^2v_T^2.
\label{}
\end{align}

Similarly to the most supersymmetry models, those LFV processes originate from the off-diagonal entries of the soft breaking terms $m_{l}^{2}$ and $m_{r}^{2}$ in the MRSSM, which are parametrized by the mass insertion
\begin{align}
(m^{2}_{l})_{IJ}=\delta ^{IJ}_{l}\sqrt{(m^{2}_{l})_{II}(m^{2}_{l})_{JJ}},
(m^{2}_{r})_{IJ}=\delta^{IJ}_{r}\sqrt{(m^{2}_{r})_{II}(m^{2}_{r})_{JJ}},\label{MI}
\end{align}
where $I,J=1,2,3$. To decrease the number of free parameters involved in our calculation, we assume that the off-diagonal entries of $m_{l}^{2}$ and $m_{r}^{2}$ in Eq.(\ref{MI}) are equal, i.e., $\delta ^{IJ}_{l}$ = $\delta ^{IJ}_{r}$ = $\delta ^{IJ}$. The experimental bounds on LFV decays, such as radiative two body decays $l_2\rightarrow l_1\gamma$, leptonic three body decays $l_2\rightarrow 3l_1$ and $\mu$-e conversion in nuclei, can give strong constraints on the parameters $\delta ^{IJ}$. In the following, we will use LFV decays $l_2\rightarrow l_1\gamma$ and $l_2\rightarrow 3 l_1$ to constrain the parameters $\delta ^{IJ}$. It is noted that $\delta ^{12}$ has been set zero in following discussion since it has no effect on the prediction of BR($\tau\rightarrow P l$). Current limits on branching ratios of $l_2\rightarrow l_1\gamma$ and $l_2\rightarrow 3 l_1$ are listed in TABLE.\ref{lfvl} \cite{PDG}.
\begin{table}[h]
\caption{Current limits of $l_2\rightarrow l_1\gamma$ and $l_2\rightarrow 3 l_1$. }
\begin{tabular}{@{}cccccc@{}} \toprule
Decay&Bound&Experiment&Decay&Bound&Experiment\\
\colrule
$\mu\rightarrow e\gamma$&$4.2\times 10^{-13}$&SPEC(2016)\cite{MEG}&$\tau\rightarrow e\gamma$&$3.3\times 10^{-8}$&BABAR(2010)\cite{BABAR}\\
$\tau\rightarrow \mu\gamma$&$4.4\times 10^{-8}$&BABAR(2010)\cite{BABAR}&$\mu\rightarrow 3e$&$1.0\times 10^{-12}$&SPEC(1988)\cite{SINDRUM}\\
$\tau\rightarrow 3e$&$2.7\times 10^{-8}$&BELL(2010)\cite{BELL}&$\tau\rightarrow 3\mu$&$2.1\times 10^{-8}$&BELL(2010)\cite{BELL}\\\botrule
\end{tabular}
\label{lfvl}
\end{table}

\begin{figure}[htbp]
\setlength{\unitlength}{1mm}
\centering
\begin{minipage}[c]{0.96\columnwidth}
\includegraphics[width=0.5\columnwidth]{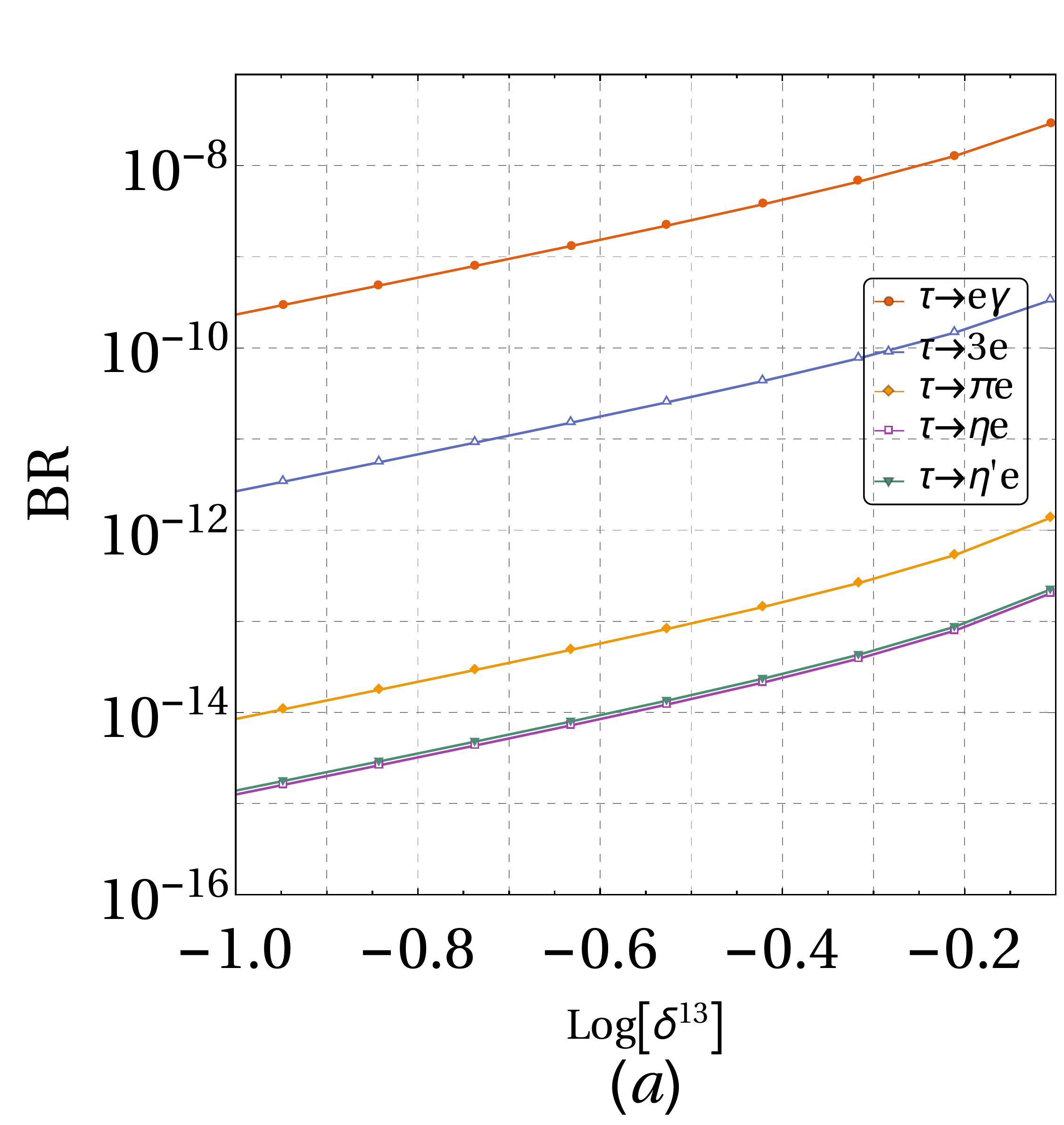}%
\includegraphics[width=0.5\columnwidth]{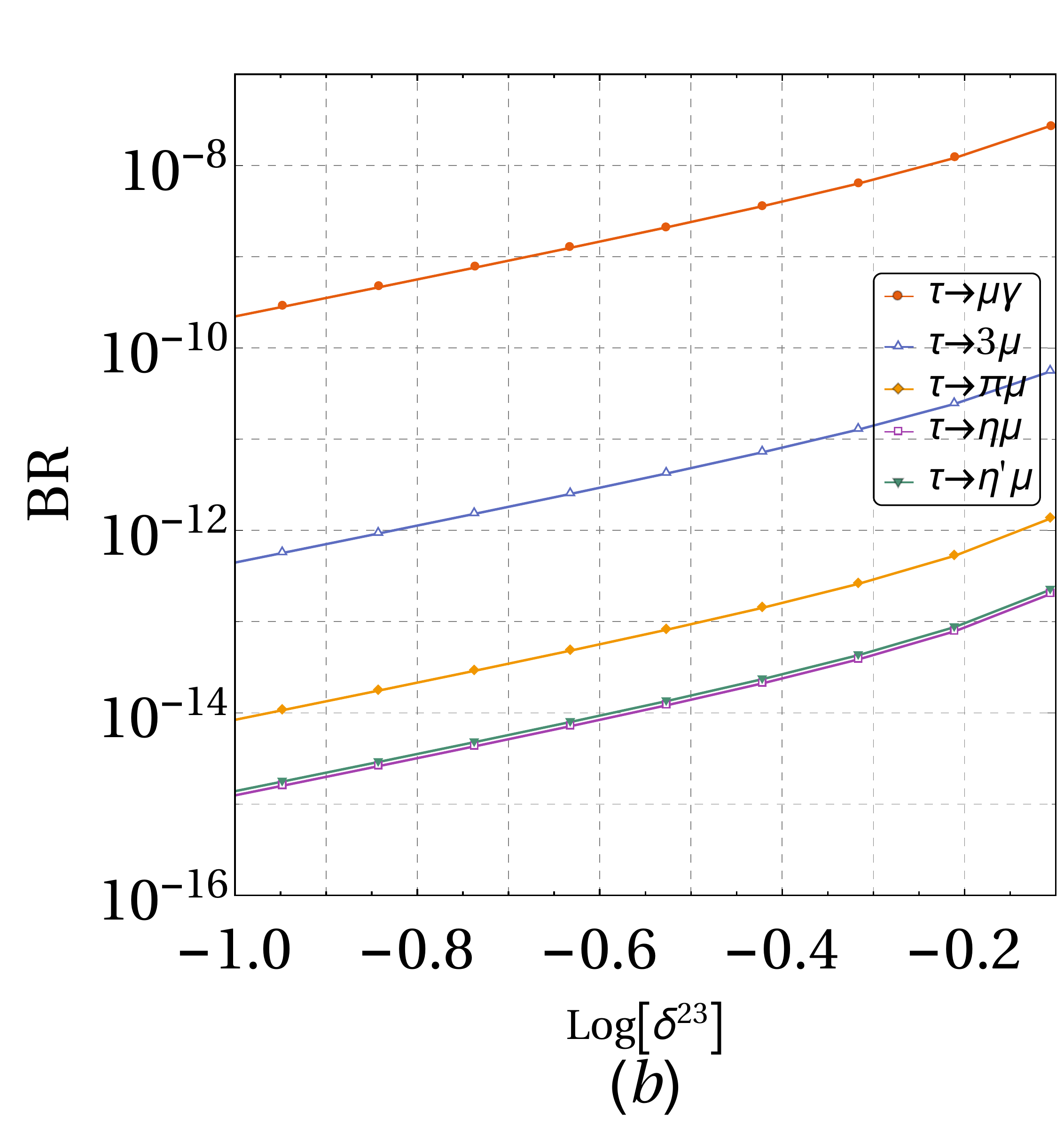}
\end{minipage}
\caption[]{Left panel: Dependence on mass insertion $\delta^{13}$ of BR($\tau\rightarrow P e$). Right panel: Dependence on mass insertion $\delta^{23}$ of BR($\tau\rightarrow P \mu$).}
\label{figD}
\end{figure}

Taking $\delta^{23}=0$ and the parameters in Eq.(\ref{N1}), we plot the predictions of BR$(\tau\rightarrow P e)$ versus Log[$\delta^{13}$] in the left panel of FIG.\ref{figD}. Taking $\delta^{13}=0$ and the parameters in Eq.(\ref{N1}), we plot the predictions of BR$(\tau\rightarrow P \mu)$ versus Log[$\delta^{23}$] in the right panel of FIG.\ref{figD}. A linear relationship in logarithmic scale is displayed between BR$(\tau\rightarrow P e(\mu))$ and the flavor violating parameter $\delta^{13}$($\delta^{23}$). The actual dependence on $\delta^{13}$ or $\delta^{23}$ is quadratic. The mentioned linear dependence is due to the fact that both x axis and y axis in FIG.\ref{figD} are logarithmically scaled. In FIG.\ref{figD} the following hierarchy is shown, BR$(\tau\rightarrow \pi e)>$BR$(\tau\rightarrow \eta' e)>$BR$(\tau\rightarrow \eta e)$ and BR$(\tau\rightarrow \pi \mu)>$BR$(\tau\rightarrow \eta' \mu)>$BR$(\tau\rightarrow \eta \mu)$. The predictions on BR$(\tau\rightarrow \eta e(\mu))$ and BR$(\tau\rightarrow \eta' e(\mu))$ are very close to each other. At $\delta^{13}(\delta^{23})$ = $10^{-0.25}\sim0.56$, the prediction on BR$(\tau\rightarrow e(\mu)\gamma)$ is around $10^{-8}$ and this is very close to the current experimental bound. The prediction on BR$(\tau\rightarrow 3e(\mu))$ is around $10^{-10}$ and this is two orders of magnitude below the current experimental bound. The decay channels $\tau\rightarrow e(\mu)\gamma$ can set more strong constraint than channels $\tau\rightarrow 3e(\mu)$ on the flavor violating parameters. The predictions on BR$(\tau\rightarrow P l)$ are far below the current experimental bounds. The prediction on BR$(\tau\rightarrow \pi e(\mu))$ is around $10^{-13}$ and this is five orders of magnitude below the current experimental bound and three orders of magnitude below the future experimental sensitivity.

\begin{figure}[htbp]
\setlength{\unitlength}{1mm}
\centering
\begin{minipage}[c]{0.96\columnwidth}
\includegraphics[width=0.5\columnwidth]{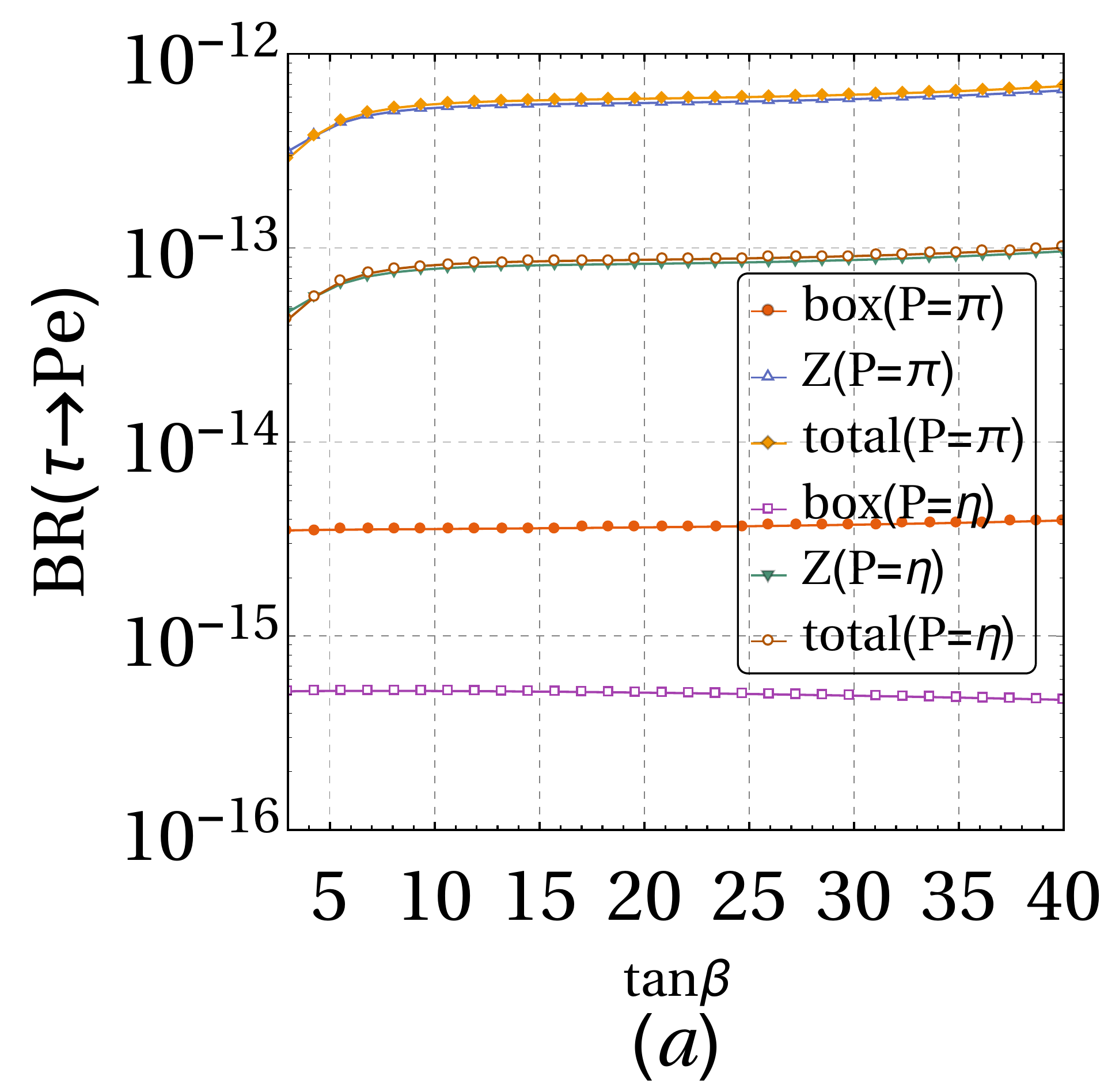}%
\includegraphics[width=0.5\columnwidth]{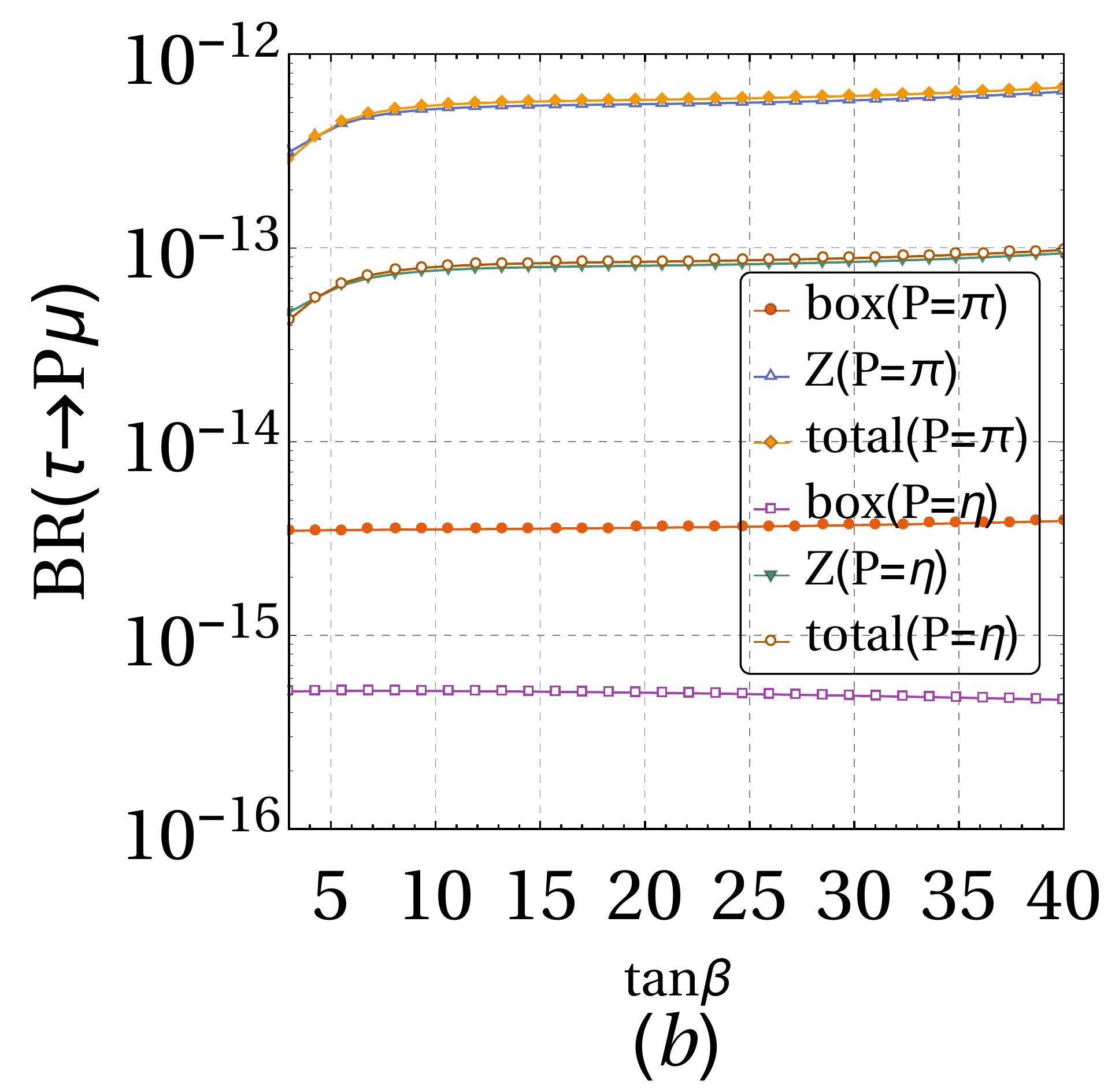}
\end{minipage}
\caption[]{Contributions to BR($\tau\rightarrow P e$) and BR($\tau\rightarrow P \mu$) from $Z$ penguins, box diagrams and total diagrams. BR($\tau\rightarrow \eta' e(\mu)$) are not shown in plots cause they are very close to BR($\tau\rightarrow \eta e(\mu)$).}
\label{figHZBT}
\end{figure}

Taking $\delta^{13}=0.5$, $\delta^{23}=0$ and the parameters in Eq.(\ref{N1}), we plot the predictions of BR$(\tau\rightarrow P e)$ from various parts as a function of $\tan \beta$ in the left panel of FIG.\ref{figHZBT}. Taking $\delta^{13}=0$, $\delta^{23}=0.5$ and the parameters in Eq.(\ref{N1}), we plot the theoretical predictions of BR$(\tau\rightarrow P\mu)$ from various parts as a function of $\tan \beta$ in the right panel of FIG.\ref{figHZBT}. The lines corresponding to $Z$ penguin and box diagram indicate the values of BR$(\tau\rightarrow P l)$ are given by only the listed contribution with all others set to zero. The total prediction for BR$(\tau\rightarrow P l)$ is also indicated. We observe that $Z$ penguin dominates the prediction on BR$(\tau\rightarrow Pl)$ in a large region of the parameter space. For larger values of $\tan\beta$, the prediction from $Z$ penguin changes slowly, since it is directly proportional to $\frac{1}{1+\tan^2\beta}$ \cite{Fok}. The contribution from box diagram is not sensitive to $\tan\beta$ and at least two orders of magnitude below $Z$ penguin. This is because of the small couplings of neutralino/chargino to quark and squark in box diagram.

As mentioned above, the contribution from Higgs penguin is negligible due to the small couplings of the scalar and pseudo-scalar Higgs to the light quarks. However, if the Higgs couplings to the strange components of the $\eta$ and $\eta'$ mesons are large enough, which result in large $A^0$-$\eta$ and $A^0$-$\eta'$ mixing, Higgs-mediated contribution could dominate the predictions on $\tau\rightarrow \mu\eta(\eta')$ \cite{Arganda}. Furthermore, it was pointed out in Ref.\cite{Celis1} that a one loop Higgs generated gluon operator does not suffer of the light-quark mass suppression and could give a sizeable contribution.
The Wilson coefficients $C^S_{XY}$ and $B^S_{XY}$ corresponding to the Higgs penguin in Eq.(\ref{BC}) can also contribute to the LFV process $\mu$-e conversion. In simple formulas, the branching ratios BR($\tau\rightarrow P l$) and conversion rate CR($\mu$-e, nucleus) are given by
\begin{eqnarray}
BR(\tau\rightarrow P l)&\sim&\frac{m^5_{\tau}f^2_{P}}{64\pi}|C^S_{XY}|^2\sim 10^{-4}|C^S_{XY}|^2,\nonumber\\
CR(\mu-e, \textup{nucleus})&\sim& \frac{3Zm^5_{\mu}\alpha^3Z^4_{eff}F^2_p}{8\pi^2\Gamma_{capt}}|C^S_{XY}|^2\sim 10^{10}|C^S_{XY}|^2(\textup{nucleus = Ti}).\nonumber
\end{eqnarray}
Thus, the predicted CR($\mu$-e, nucleus) from Higgs penguin is much larger than BR($\tau\rightarrow P l$) though both are negligible compared to the total contribution.

\begin{figure}[htbp]
\setlength{\unitlength}{1mm}
\centering
\begin{minipage}[c]{0.96\columnwidth}
\includegraphics[width=0.5\columnwidth]{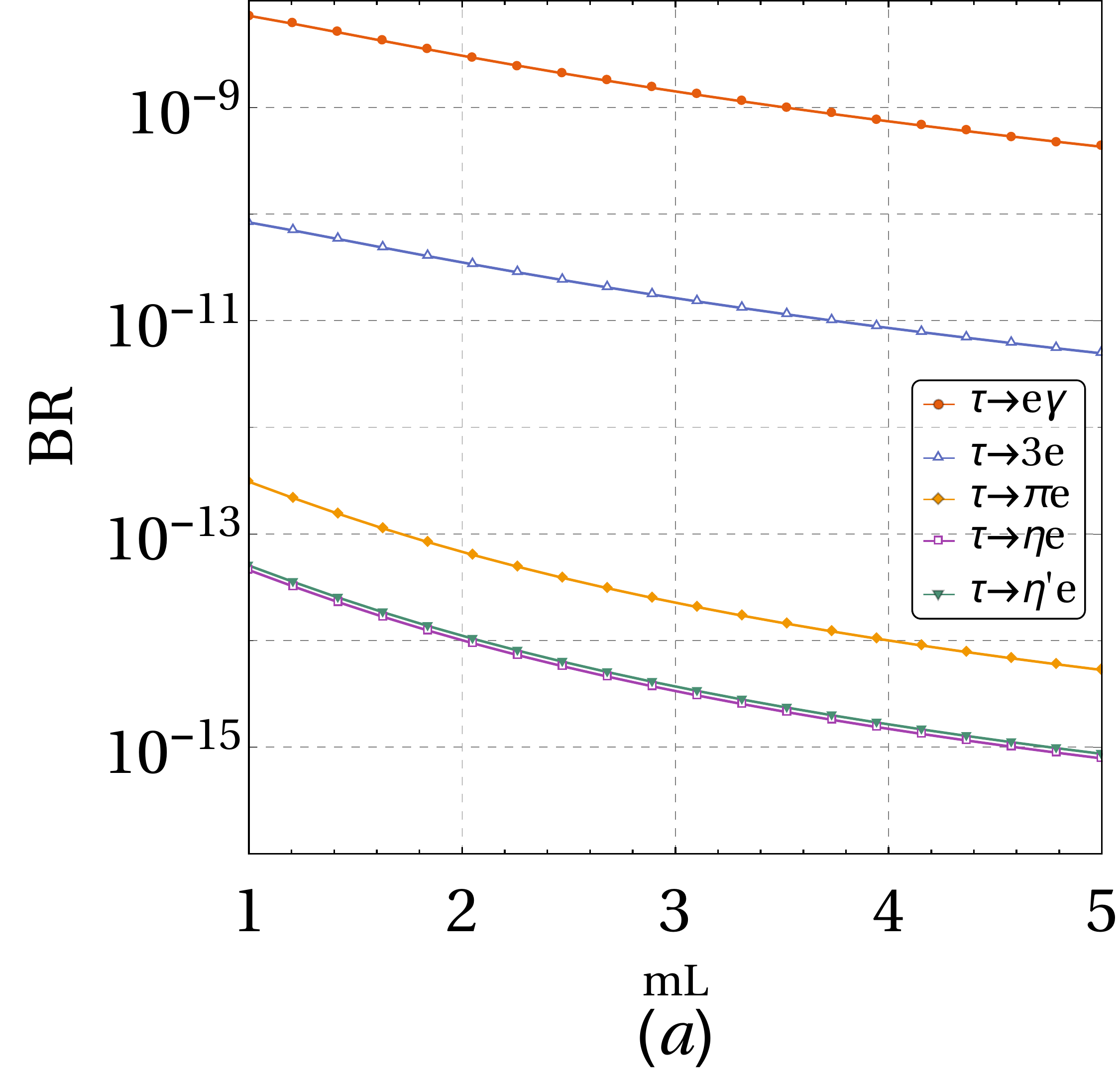}%
\includegraphics[width=0.5\columnwidth]{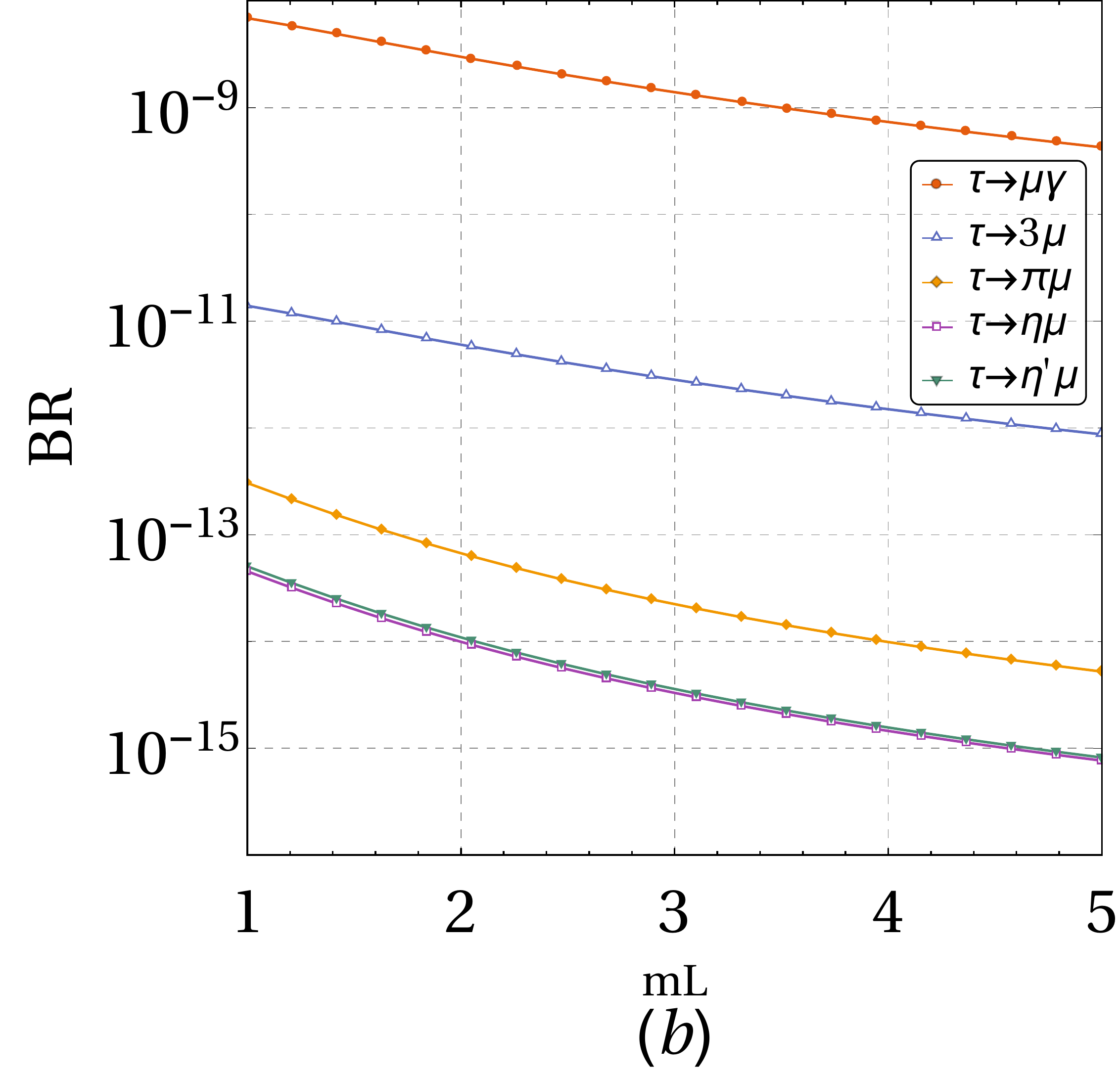}
\end{minipage}
\caption[]{Left panel: Dependence on $mL$ of BR($\tau\rightarrow P e$). Right panel: Dependence on $mL$ of BR($\tau\rightarrow P \mu$). The mass parameter $mL$ is in TeV. }
\label{figmL}
\end{figure}

Taking $\delta^{13}=0.5$, $\delta^{23}=0$ and the parameters in Eq.(\ref{N1}), we plot the predictions of BR$(\tau\rightarrow P e)$ as a function of the diagonal entries $mL$ in the left panel of FIG.\ref{figmL}. Taking $\delta^{13}=0$, $\delta^{23}=0.5$ and parameters in Eq.(\ref{N1}), we plot the predictions of BR$(\tau\rightarrow P \mu)$ as a function of the diagonal entries $mL$ in the right panel of FIG.\ref{figmL}. Here, $mL$$=\sqrt{(m^2_l)_{11}}=\sqrt{(m^2_l)_{22}}=\sqrt{(m^2_l)_{33}}=\sqrt{(m^2_r)_{11}}
=\sqrt{(m^2_r)_{22}}=\sqrt{(m^2_r)_{33}}$. The prediction on BR$(\tau\rightarrow l\gamma)$ is at the level of $\mathcal{O}(10^{-9})$ and this is very close to the future experimental sensitivities \cite{Aushev}. The prediction on BR$(\tau\rightarrow 3l)$ is at the level of $\mathcal{O}(10^{-11})$ and this is two orders of magnitude below the future experimental sensitivities \cite{Aushev}. The predictions on BR$(\tau\rightarrow l\gamma)$, BR$(\tau\rightarrow 3l)$ and BR$(\tau\rightarrow P l)$ in the MRSSM decreas as the slepton mass parameter $mL$ varies from 1 TeV to 5 TeV.

\begin{figure}[htbp]
\setlength{\unitlength}{1mm}
\centering
\begin{minipage}[c]{0.96\columnwidth}
\includegraphics[width=0.5\columnwidth]{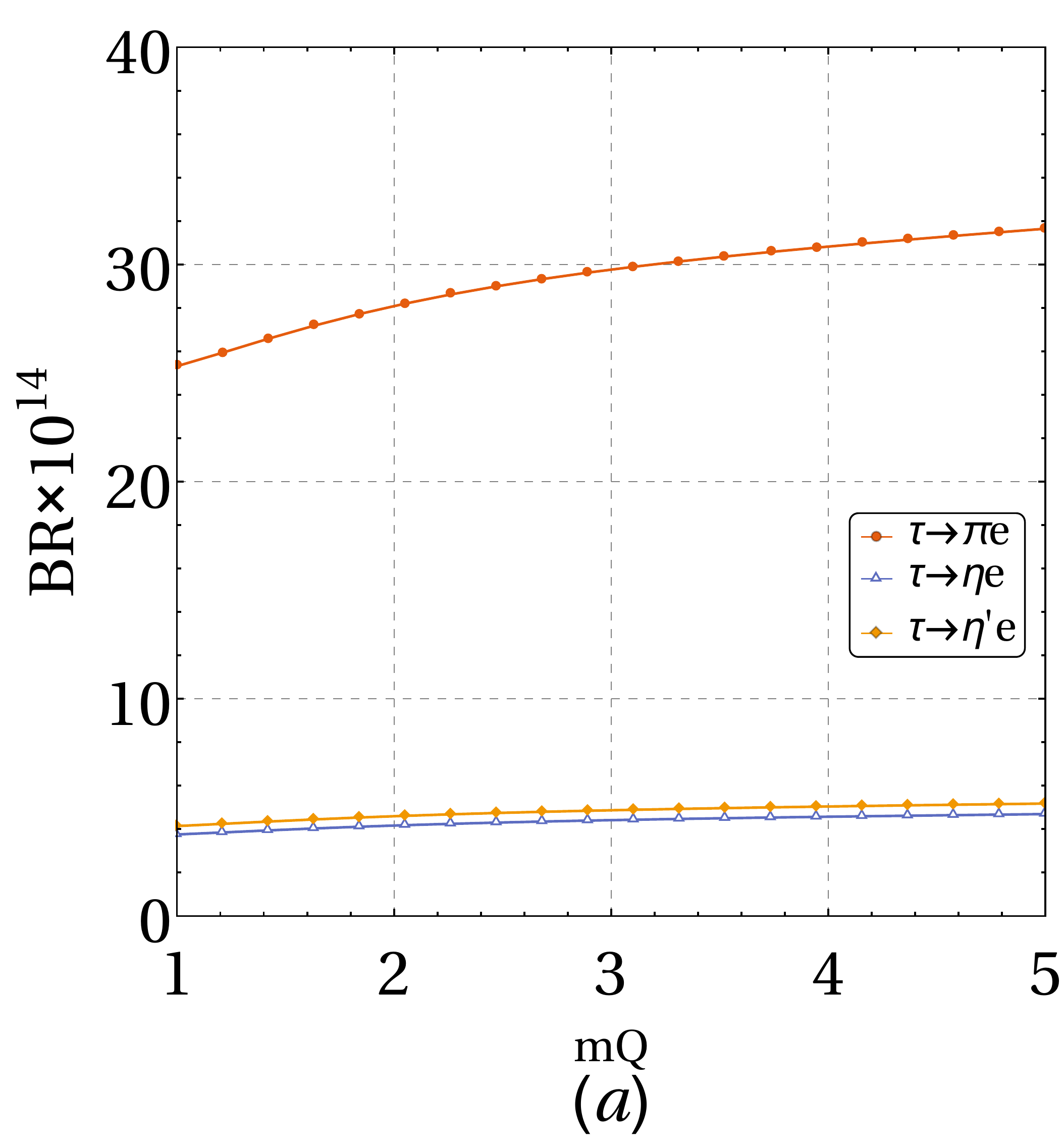}%
\includegraphics[width=0.5\columnwidth]{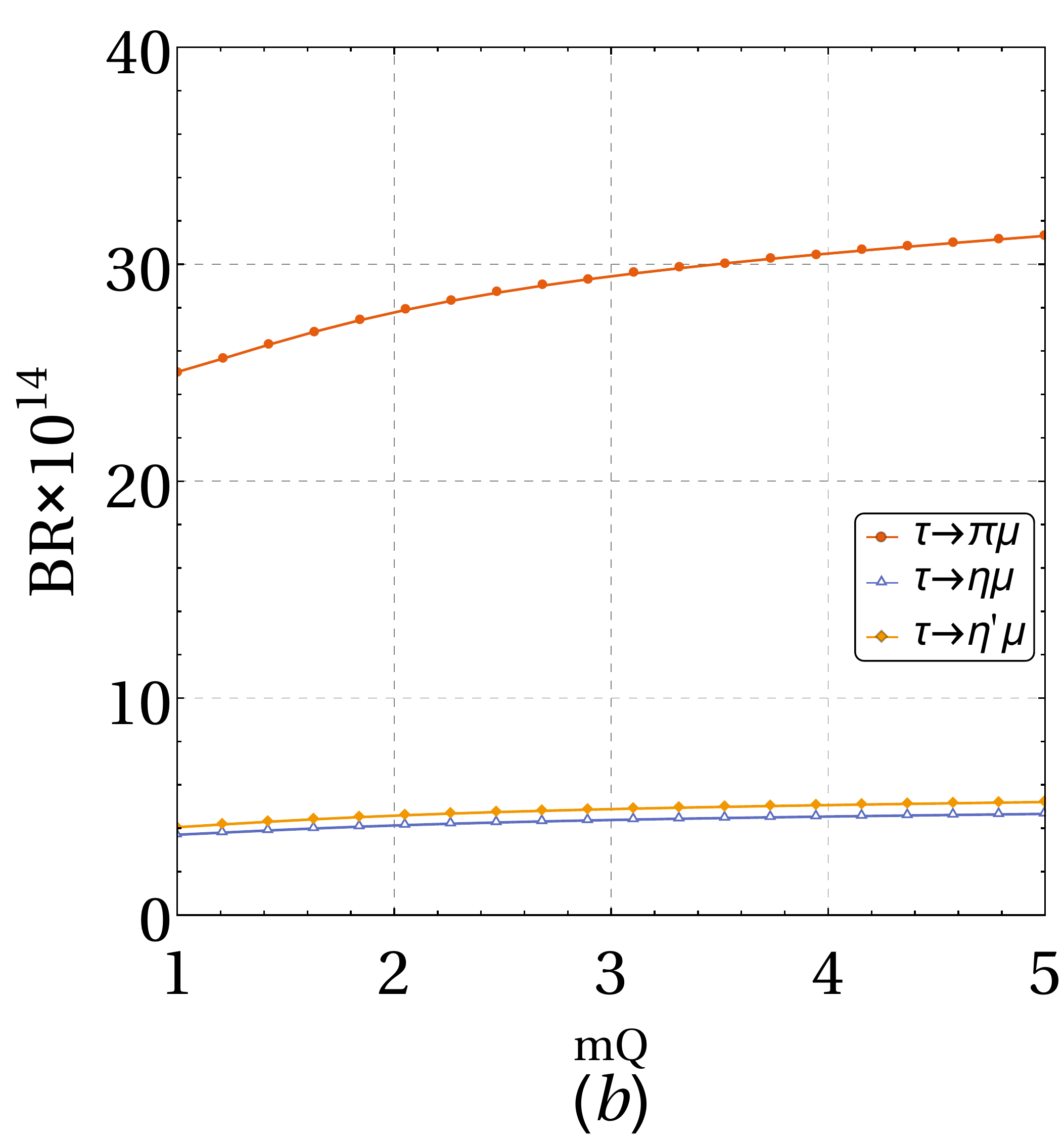}
\end{minipage}
\caption[]{Left panel: Dependence on $mQ$ of BR($\tau\rightarrow P e$). Right panel: Dependence on $mQ$ of BR($\tau\rightarrow P \mu$). Mass parameter $mQ$ is in TeV.}
\label{figmQ}
\end{figure}

Taking $\delta^{13}=0.5$, $\delta^{23}=0$ and the parameters in Eq.(\ref{N1}), we plot the predictions of BR $(\tau\rightarrow Pe)$ as a function of the squark mass parameter $mQ$ in the left panel of FIG.\ref{figmQ}. Taking $\delta^{13}=0$, $\delta^{23}=0.5$ and the parameters in Eq.(\ref{N1}), we plot the predictions of BR $(\tau\rightarrow P\mu)$ as a function of the squark mass parameter $mQ$ in the right panel of FIG.\ref{figmQ}. Here, $mQ$=$\sqrt{(m^2_{\tilde{q}})_{11}}=\sqrt{(m^2_{\tilde{u}})_{11}}=\sqrt{(m^2_{\tilde{d}})_{11}}=\sqrt{(m^2_{\tilde{q}})_{22}}
=\sqrt{(m^2_{\tilde{u}})_{22}}=\sqrt{(m^2_{\tilde{d}})_{22}}=\sqrt{(m^2_{\tilde{q}})_{33}}=\sqrt{(m^2_{\tilde{u}})_{33}}=\sqrt{(m^2_{\tilde{d}})_{33}}$.  We clearly see that both the predictions on BR($\tau\rightarrow P e$) and BR($\tau\rightarrow P \mu$), which increase slowly as $mQ$ varies from 1 TeV to 5 TeV, are not sensitive to $mQ$. The off-diagonal entries $\delta^{IJ}_{\tilde{q},\tilde{u},\tilde{d}}$ of the squark mass matrices $m^2_{\tilde{q}}$, $m^2_{\tilde{u}}$ and $m^2_{\tilde{d}}$ could give additional contributions to BR($\tau\rightarrow P l$). Taking into account the experimental constraints on $\delta^{IJ}_{\tilde{q},\tilde{u},\tilde{d}}$ from low energy B meson physics observables, such as, BR($\bar{B}\rightarrow X_s\gamma$), BR($B^0_{s,d}\rightarrow \mu^+\mu^-$), the prediction of BR($\tau\rightarrow P l$) takes values along a narrow band. Thus, the prediction of BR$(\tau\rightarrow P l)$ is also not sensitive to the off-diagonal entries of the squark mass matrices.

We are also interested to the effect from other parameters on the prediction of BR($\tau\rightarrow P l$) in the MRSSM. The predicted BR($\tau\rightarrow P l$) decreases slowly along with the increase of the wino-triplino mass parameter $M_D^W$. However, the valid region of $M_D^W$ is constrained by the boundary conditions at the unification scale, and unphysical masses of neutral Higgs and charged Higgs are obtained when $M_D^W$ above 1 TeV. By scanning over these parameters, which are shown in Eq.(\ref{N2}),
\begin{equation}
\begin{array}{l}
-1.5<\lambda_d, \lambda_u, \Lambda_d, \Lambda_u<1.5,\\
300\text{ GeV}<\mu_d, \mu_u, m_A<1000\text{ GeV},
\end{array}\label{N2}
\end{equation}
the prediction is shown in relation to one input parameter (e.g. $\lambda_d$ or others).
The parameters $\lambda_d, \lambda_u, \Lambda_u$ are constrained in a narrow region by the boundary conditions which is close to the benchmark points and $\Lambda_d$ can vary in the whole region. The results show that varying those parameters in Eq.(\ref{N2}) have very small effect on the prediction of BR($\tau\rightarrow P l$) which takes values along a narrow band.

\section{Conclusions\label{sec4}}
In this work, taking into account the constraints from LFV decays $\tau\rightarrow e (\mu) \gamma$ and $\tau\rightarrow 3 e (\mu)$ on the flavor violating parameters, we analyze the LFV decays $\tau\rightarrow P l$ in the framework of the minimal R-symmetric supersymmetric standard model.

We observe that $Z$ penguin dominates the prediction on BR$(\tau\rightarrow P e (\mu))$, and other contributions are less dominant or negligible. As we all konw the LFV processes $l_2\rightarrow l_1 \gamma$ can only obtain the dipole contribution from the gamma penguin. The LFV processes $l_2\rightarrow3 l_1$ and $\mu$-e conversion can obtain contributions from the gamma penguin (including dipole and non-dipole), $Z$ penguin, Higgs penguin and box diagram. While, for the LFV processes $\tau\rightarrow Pl$, contributions from $Z$ penguin, Higgs penguin and box diagram are included except for the gamma penguin. It is interesting to consider if it is possible to find a parameter region that LFV processes $\tau\rightarrow e(\mu)\gamma$ and $\tau$-$e(\mu)$ conversion are not excluded and which could still give large contribution to $\tau\rightarrow Pl$. The answer is negative at least in this work. The prediction of BR$(\tau\rightarrow P l)$ is two or three orders of magnitude below BR$(\tau\rightarrow 3l)$.

In the MRSSM, the prediction on BR$(\tau\rightarrow P e)$ and BR$(\tau\rightarrow P \mu)$ is affected by the mass insertions $\delta^{13}$ and $\delta^{23}$, respectively. The prediction on BR$(\tau\rightarrow P e)$ would be zero if $\delta^{13}$=0 is assumed, and so are the prediction on BR$(\tau\rightarrow P \mu)$ if $\delta^{23}$=0 is assumed. Taking into account the experimental bounds on BR($\tau\rightarrow e(\mu)\gamma$) and BR($\tau\rightarrow 3e(\mu)$), the values of $\delta^{13}$ and $\delta^{23}$ are constrained around 0.5. The predictions on BR$(\tau\rightarrow P e)$ and BR$(\tau\rightarrow P \mu)$ are found to be at the level of $\mathcal{O}(10^{-13}$-$10^{-14})$, which are five orders of magnitude below the present experimental upper limits. The processes $\tau\rightarrow  \pi l$ may be the most competitive LFV semileptonic tau decay channels. The future prospects of BR$(\tau\rightarrow Pl)$ in Belle II are extrapolated at the level of $\mathcal{O}(10^{-9}$-$10^{-10})$ \cite{Altmannshofer} and they are three orders of magnitude above the predictions in the MRSSM. Thus, LFV decays $\tau\rightarrow Pl$ may be out reach of the near future experiments.

\appendix

\section{Mass matrices at tree level in the MRSSM\label{secapp}}

In the weak basis $(\phi_d,\phi_u,\phi_S,\phi_T)$, the scalar Higgs boson mass matrix and the diagonalization procedure are
\begin{eqnarray}
{\cal M}_h &=& \left(
\begin{array}{cc}
{\cal M}_{11}&{\cal M}_{21}^T\\
{\cal M}_{21}&{\cal M}_{22}\\
\end{array}
\right), Z^h {\cal M}_{h} (Z^{h})^{\dagger}={\cal M}_{h}^{\textup{diag}},
\end{eqnarray}
where the submatrices ($c_{\beta}=cos\beta$, $s_{\beta}=sin\beta$) are
\begin{eqnarray}
{\cal M}_{11}&=& \left(
\begin{array}{cc}
m_Z^2c^2_{\beta}+m_A^2s^2_{\beta}&-(m_Z^2+m_A^2)s_{\beta}c_{\beta}\\
-(m_Z^2+m_A^2)s_{\beta}c_{\beta}&m_Z^2s^2_{\beta}+m_A^2c^2_{\beta}\\
\end{array}
\right),\nonumber\\
{\cal M}_{21}&=& \left(
\begin{array}{cc}
v_d(\sqrt{2}\lambda_d\mu_d^{eff,+}-g_1M_B^D)&
v_u(\sqrt{2}\lambda_u\mu_u^{eff,-}+g_1M_B^D) \\
v_d(\Lambda_d\mu_d^{eff,+}+g_2M_W^D)& -
v_u(\Lambda_u\mu_u^{eff,1}+g_2M_W^D) \\
\end{array}
\right),\nonumber\\
{\cal M}_{22}&=& \left(
\begin{array}{cc}
4 (M_B^D)^2+m_S^2+\frac{\lambda_d^2 v_d^2+\lambda_u^2 v_u^2}{2} \;
& \frac{\lambda_d \Lambda_d v_d^2-\lambda_u \Lambda_u v_u^2}{2 \sqrt{2}} \\
 \frac{\lambda_d \Lambda_d v_d^2-\lambda_u \Lambda_u v_u^2}{2 \sqrt{2}} \;
 & 4 (M_W^D)^2+m_T^2+\frac{\Lambda_d^2 v_d^2+\Lambda_u^2 v_u^2}{4}\\
\end{array}
\right).\nonumber
\end{eqnarray}

In the weak basis $(\sigma_d,\sigma_u,\sigma_S,\sigma_T)$, the pseudo-scalar Higgs boson mass matrix and the diagonalization procedure are
\begin{eqnarray}
{\cal M}_{A^0} &=& \left(
\begin{array}{cccc}
B_{\mu} \frac{v_u}{v_d} & B_{\mu} & 0 & 0 \\
B_{\mu} &  B_{\mu} \frac{v_d}{v_u} & 0 & 0 \\
0 & 0 & m_S^2+\frac{\lambda_d^2 v_d^2+\lambda_u^2 v_u^2 }{2} & \frac{\lambda_d\Lambda_d v_d^2-\lambda_u\Lambda_u v_u^2}{2 \sqrt{2}} \\
0 & 0 & \frac{\lambda_d\Lambda_d v_d^2-\lambda_u\Lambda_u v_u^2}{2 \sqrt{2}}& m_T^2+ \frac{\Lambda_d^2 v_d^2+\Lambda_u^2 v_u^2 }{4}\\
\end{array}
\right), Z^A {\cal M}_{A^0} (Z^{A})^{\dagger}={\cal M}_{A^0}^{\textup{diag}} .
\end{eqnarray}

In the weak basis of four neutral electroweak two-component fermions $\xi_i$=($\tilde{B}$,$\tilde{W}^0$,$\tilde{R}^0_d$,$\tilde{R}^0_u$) with R-charge 1 and four neutral electroweak two-component fermions $\varsigma_i$=($\tilde{S}$,$\tilde{T}^0$,$\tilde{H}^0_d$,$\tilde{H}^0_u$) with R-charge -1, the neutralino mass matrix and the diagonalization procedure are
\begin{eqnarray}
m_{\chi^0} &=& \left(
\begin{array}{cccc}
M^{B}_D &0 &-\frac{1}{2} g_1 v_d  &\frac{1}{2} g_1 v_u \\
0 &M^{W}_D &\frac{1}{2} g_2 v_d  &-\frac{1}{2} g_2 v_u \\
- \frac{1}{\sqrt{2}} \lambda_d v_d  &-\frac{1}{2} \Lambda_d v_d  &-\mu_d^{eff,+}&0\\
\frac{1}{\sqrt{2}} \lambda_u v_u  &-\frac{1}{2} \Lambda_u v_u  &0 &\mu_u^{eff,-}\end{array}
\right),(N^{1})^{\ast} m_{\chi^0} (N^{2})^{\dagger} =m_{\chi^0}^{\textup{diag}}.
\end{eqnarray}
The mass eigenstates $\kappa_i$ and $\varphi_i$, and physical four-component Dirac neutralinos are
\begin{equation}
\xi_i=\sum^4_{j=1}(N^1_{ji})^{\ast}\kappa_j, \varsigma_i=\sum^4_{j=1}(N^2_{ij})^{\ast}\varphi_j, \chi^0_i=\left(
\begin{array}{c}
\kappa_i\\
\varphi_i^{\ast}\\
\end{array}
\right).\nonumber
\end{equation}

In the basis $\xi^+_i$=($\tilde{W}^+$, $\tilde{R}^+_d$) and $\varsigma^-_i$=($\tilde{T}^-$, $\tilde{H}^-_d$), the $\chi^{\pm}$-charginos mass matrix and the diagonalization procedure are
\begin{equation}
m_{\chi^{\pm}} = \left(
\begin{array}{cc}
g_2 v_T  + M^{W}_D &\frac{1}{\sqrt{2}} \Lambda_d v_d \\
\frac{1}{\sqrt{2}} g_2 v_d  &-\frac{1}{2} \Lambda_d v_T  + \frac{1}{\sqrt{2}} \lambda_d v_S  + \mu_d\end{array}
\right),(U^{1})^{\ast} m_{\chi^{\pm}} (V^{1})^{\dagger} =m_{\chi^{\pm}}^{\textup{diag}}.
\end{equation}
The mass eigenstates $\lambda^{\pm}_i$ and physical four-component Dirac charginos are
\begin{equation}
\xi^+_i=\sum^2_{j=1}(V^1_{ij})^{\ast}\lambda^+_j, \varsigma^-_i=\sum^2_{j=1}(U^1_{ji})^{\ast}\lambda^-_j, \chi^{\pm}_i=\left(
\begin{array}{c}
\lambda^+_i\\
\lambda_i^{-\ast}\\
\end{array}
\right).\nonumber
\end{equation}

The mass matrix for up squarks and down squarks, and the relevant diagonalization procedure are
\begin{equation}
\begin{array}{l}
m^2_{\tilde{u}} = \left(
\begin{array}{cc}
(m^2_{\tilde{u}})_{LL} &0 \\
0  &(m^2_{\tilde{u}})_{RR}\end{array}
\right), Z^U m^2_{\tilde{u}} (Z^{U})^{\dagger} =m^{2,\textup{diag}}_{\tilde{u}}, \\
m^2_{\tilde{d}} = \left(
\begin{array}{cc}
(m^2_{\tilde{d}})_{LL} &0 \\
0  &(m^2_{\tilde{d}})_{RR}\end{array}
\right),Z^D m^2_{\tilde{d}} (Z^{D})^{\dagger} =m^{2,\textup{diag}}_{\tilde{d}},\label{sud}
\end{array}
\end{equation}
where
\begin{align}
(m^2_{\tilde{u}})_{LL} &=m_{\tilde{q}}^2+ \frac{1}{2} v_{u}^{2} |Y_{u}|^2
+\frac{1}{24}(g_1^2-3g_2^2)(v_{u}^{2}- v_{d}^{2}) +\frac{1}{3}g_1 v_S M_D^B+g_2v_TM_D^W ,\nonumber\\
(m^2_{\tilde{u}})_{RR} &= m_{\tilde{u}}^2+\frac{1}{2}v_u^2|Y_u|^2+\frac{1}{6}g_1^2( v_{d}^{2}- v_{u}^{2})-\frac{4}{3} g_1v_SM_D^B,\nonumber\\
(m^2_{\tilde{d}})_{LL} &=m_{\tilde{q}}^2+ \frac{1}{2} v_{d}^{2} |Y_{d}|^2
+\frac{1}{24}(g_1^2+3g_2^2)(v_{u}^{2}- v_{d}^{2}) +\frac{1}{3}g_1 v_S M_D^B-g_2v_TM_D^W ,\nonumber\\
(m^2_{\tilde{d}})_{RR} &= m_{\tilde{d}}^2+\frac{1}{2}v_d^2|Y_d|^2+\frac{1}{12}g_1^2( v_{u}^{2}- v_{d}^{2})+\frac{2}{3} g_1v_SM_D^B.\nonumber
\end{align}

\begin{acknowledgments}
\indent\indent
This work has been supported partly by the National Natural Science Foundation of China (NNSFC) under Grants No.11905002 and No.11805140, the Scientific Research Foundation of the Higher Education Institutions of Hebei Province under Grant No. BJ2019210, the Foundation of Baoding University under Grant No. 2018Z01, the Foundation of Department of Education of Liaoning province under Grant No. 2020LNQN14, and the Natural Science Foundation of Hebei province under Grant No. A2020201002.

\end{acknowledgments}


\begin{thebibliography}{99}

\bibitem{Cei}
F. Cei and D. Nicolo, Adv. High Energy Phys. 2014, 282915(2014).
\bibitem{Hayasaka}
K. Hayasaka and J. Phys. Conf. Ser. 171, 012079(2009).
\bibitem{Belle}
Belle, BABAR Collaboration, G. Vasseur, PoS Beauty2014, 041(2015).
\bibitem{Belle1}
Belle Collaboration, Y. Miyazaki et al., Phys. Lett. B 719, 346(2013).
\bibitem{Belle2}
Belle-II Collaboration, B. A. Shwartz, Nucl. Part. Phys. Proc. 260, 233(2015).

\bibitem{PDG}
M. Tanabashi et al., (Particle Data Group), Phys. Rev. D 98, 030001(2018).
\bibitem{BELLEt}
Y. Miyazaki et al., Belle collaboration, Phys. Lett. B 648, 341(2007).
\bibitem{BABARt}
B. Aubert et al., BABAR collaboration, Phys. Rev. Lett. 98, 061803(2007).
\bibitem{Altmannshofer}
W. Altmannshofer et al. (Belle II Collaboration), arXiv:1808.10567.
\bibitem{Li}
W. J. Li, Y. D. Yang and X. D. Zhang, Phys. Rev. D 73, 073005(2006).
\bibitem{Kanemura}
S. Kanemura, T. Ota and K. Tsumura, Phys. Rev. D 73, 016006(2006).
\bibitem{hua}
T. Hua and C.-X. Yue, Commun. Theor. Phys. 62 (2014) 3, 388-392.
\bibitem{yue}
C.-X. Yue, L.-H. Wang and W. Ma, Phys. Rev. D 74, 115018(2006).
\bibitem{Goto}
T. Goto, Y. Okada and Y. Yamamoto, Phys. Rev. D 83, 053011(2011).
\bibitem{Lami}
A. Lami, J. Portoles and P. Roig, Phys. Rev. D 93, 076008(2016).
\bibitem{Carpentier}
M. Carpentier and S. Davidson, Eur. Phys. J. C 70, 1071(2010).
\bibitem{Dorsner}
I, Dorsner, J. Drobnak, S. Fajfer, J. F. Kamenik and N. Kosnik, J. High Energy Phys. 1111, 002(2011).
\bibitem{LiZH}
Z.-H. Li, Y. Li and H.-X. Xu, Phys. Lett. B 677, 150(2009).
\bibitem{Ilakovac}
A. Ilakovac, Bernd A. Kniehl and A. Pilaftsis, Phys. Rev. D 52, 3993(1995).
\bibitem{Garcia}
M. C. Gonzalez-Garcia and J. W. F. Valle, Mod. Phys. Lett. A 7, 477(1992).
\bibitem{Ilakovac3}
A. Ilakovac, Phys. Rev. D 62, 036010(2000).
\bibitem{He1}
X.-G. He and S. Oh, J. High Energy Phys. 0909, 027(2009).
\bibitem{Arhrib}
A. Arhrib, R. Benbrik and C.-H. Chen, Phys. Rev. D 81, 113003(2010).
\bibitem{Sher}
M. Sher, Phys. Rev. D 66, 057301(2002).
\bibitem{Fukuyama}
T. Fukuyama, A. Ilakovac and T. Kikuchi, Eur. Phys. J. C 56, 125(2008).
\bibitem{brignole}
A. Brignole and A. Rossi, Nucl. Phys. B 701, 3(2004).
\bibitem{Chen}
C.-H. Chen and C.-Q. Geng, Phys. Rev. D 74, 035010(2006).
\bibitem{Saha}
J. P. Saha and A. Kundu, Phys. Rev. D 66, 054021(2002).
\bibitem{Arganda}
E. Arganda, M.J. Herrero and J. Portoles, J. High Energy Phys. 0806, 079(2008).
\bibitem{Black}
D. Black, T. Han, H. J. He and M. Sher, Phys. Rev. D 66, 053002(2002).
\bibitem{cirigliano}
V. Cirigliano and B. Grinstein, Nucl. Phys. B 752, 18(2006).
\bibitem{He}
X.-G. He, J. Tandean and G. Valencia, Phys. Lett. B 797, 134842(2019).
\bibitem{Cai}
Y. Cai and M. A. Schmidt, J. High Energy Phys. 1602, 176(2016).
\bibitem{Cai1}
Y. Cai, M. A. Schmidt and G. Valencia, J. High Energy Phys. 1805, 143(2018).
\bibitem{Gabrielli}
E. Gabrielli, Phys.Rev. D 62, 055009(2000).
\bibitem{Petrov}
A. A. Petrov, D. V. Zhuridov, Phys. Rev. D 89, 033005(2014).
\bibitem{Celis}
A. Celis, V. Cirigliano and E. Passemar, Phys. Rev. D 89, 095014(2014).
\bibitem{Buchmuller}
W. Buchmuller and D. Wyler, Nucl. Phys. B 268, 621(1986).
\bibitem{Grzadkowski}
B. Grzadkowski, M. Iskrzynski and M. Misiak, J. High Energy Phys. 1010, 085(2010).

\bibitem{Kribs}
G. D. Kribs, E. Poppitz and N. Weiner, Phys. Rev. D 78, 055010(2008).
\bibitem{Die1}
P. Diessner and W. Kotlarski, PoS CORFU 2014, 079(2015).
\bibitem{Die2}
P. Diessner, J. Kalinowski, W. Kotlarski and D. St\"{o}ckinger, Adv. High Energy Phys. 2015, 760729(2015).
\bibitem{Die3}
P. Diessner, J. Kalinowski, W. Kotlarski and D. St\"{o}ckinger, J. High Energy Phys. 1412, 124(2014).
\bibitem{Die4}
P. Diessner, W. Kotlarski, S. Liebschner and D. St\"{o}ckinger, J. High Energy Phys. 1710, 142(2017).
\bibitem{Die5}
P. Diessner, J. Kalinowski, W. Kotlarski and D. St\"{o}ckinger, J. High Energy Phys. 1603, 007(2016).
\bibitem{Die6}
P. Diessner and G. Weiglein, J. High Energy Phys. 1907, 011(2019).
\bibitem{KSS}
W. Kotlarski, D. St\"{o}ckinger and H. St\"{o}ckinger-Kim, J. High Energy Phys. 1908, 082(2019).
\bibitem{Kumar}
A. Kumar, D. Tucker-Smith and N. Weiner, J. High Energy Phys. 1009, 111(2010).
\bibitem{Blechman}
A. E. Blechman, Mod. Phys. Lett. A 24, 633(2009).
\bibitem{Kribs1}
G. D. Kribs, A. Martin and T. S. Roy, J. High Energy Phys. 0906, 042(2009).
\bibitem{Frugiuele}
C. Frugiuele and T. Gregoire, Phys. Rev. D 85, 015016(2012).
\bibitem{Jan}
J. Kalinowski, Acta Phys. Polon. B 47, 203(2016).
\bibitem{Chakraborty}
S. Chakraborty, A. Chakraborty and S. Raychaudhuri, Phys. Rev. D 94, 035014(2016).
\bibitem{Braathen}
J. Braathen, M. D. Goodsell and P. Slavich, J. High Energy Phys. 1609, 045(2016).
\bibitem{Athron}
P. Athron, J.-hyeon Park, T. Steudtner, D. St\"{o}ckinger and A. Voigt, J. High Energy Phys. 1701, 079(2017).
\bibitem{Alvarado}
C. Alvarado, A. Delgado and A. Martin, Phys. Rev. D 97, 115044(2018).
\bibitem{sks1}
K.-S. Sun, J.-B. Chen, X.-Y. Yang and S.-K. Cui, Chin. Phys. C 43, 043101 (2019).
\bibitem{sks2}
K.-S. Sun, S.-K. Cui, W Li and H.-B. Zhang, Phys. Rev. D 102, 035029 (2020).
\bibitem{SARAH}
F. Staub, arXiv:0806.0538.
\bibitem{SARAH1}
F. Staub, Comput. Phys. Commun. 184, 1792(2013).
\bibitem{SARAH2}
F. Staub, Comput. Phys. Commun. 185, 1773(2014).
\bibitem{Flavor}
W. Porod and F. Staub, A. Vicente, Eur. Phys. J. C 74, 2992(2014).
\bibitem{SPheno1}
W. Porod, Comput. Phys. Commun. 153, 275(2003).
\bibitem{SPheno2}
W. Porod and F. Staub, Comput. Phys. Commun. 183, 2458(2012).

\bibitem{MEG}
A. M. Baldini, et al., (MEG Collaboration), Eur. Phys. J. C 76, 434(2016).
\bibitem{BABAR}
B. Aubert, et.al., (BABAR Collaboration), Phys. Rev. Lett. 104, 021802(2010).
\bibitem{SINDRUM}
U. Bellgardt et al., (SINDRUM Collaboration), Nucl. Phys. B 299, 1(1988).
\bibitem{BELL}
K. Hayasaka et al., Phys. Lett. B 687, 139(2010).
\bibitem{Fok}
R. Fok and G. D. Kribs, Phys. Rev. D 82, 035010(2010).
\bibitem{Celis1}
A. Celis, V. Cirigliano and E. Passemar, Phys. Rev. D 89, 013008(2014).


\bibitem{Aushev}
T. Aushev, et al., arXiv:1002.5012.

\end{thebibliography}
\end{document}